\IfFileExists{./paper_tex_template.bib}{}{\errmessage{Please create radiology.bib as a 
symbolic link, with 'ln -s /path/to/radiology-db/radiology.bib 
paper_tex_template.bib'}}

\RequirePackage[]{graphicx} % Loaded here to show images in draft mode, too

\def\mode{1} % 1 is Arxiv

\if 1\mode
	\documentclass[journal]{article}
\fi

\if 2\mode
	\documentclass[num-refs]{wiley-article_final}	
\fi

\usepackage{etoolbox}
\newtoggle{MRM}
\newtoggle{ARXIV}

\newtoggle{SUPMATERIAL}
\togglefalse{SUPMATERIAL}

\if 1\mode
	\toggletrue{ARXIV}
	\togglefalse{MRM}
\fi

\if 2\mode
	\toggletrue{MRM}
	\togglefalse{ARXIV}
\fi

\iftoggle{MRM}{
	\usepackage[nomarkers, nolists]{endfloat}
}

\usepackage[utf8]{inputenc}
\usepackage{amsmath}
\usepackage{bm}
\usepackage{bbold}
\usepackage{authblk}
\usepackage{amssymb}
\usepackage{graphicx}
\usepackage{ifdraft}
\usepackage{marginnote}
\usepackage{pdfpages}
\usepackage{tabularx,booktabs}
\usepackage{ifpdf}
\usepackage{float}

\ifpdf
	\usepackage[]{hyperref}
\else
	\usepackage[dvipdfm]{hyperref}
\fi
\usepackage{tikz}
\usepackage[capitalise]{cleveref}

\usepackage{blindtext}

\usepackage[
per-mode=fraction,
separate-uncertainty,
retain-unity-mantissa=false,
product-units=power,
table-alignment=center,
detect-all,
binary-units=true,
exponent-product=\cdot,
exponent-base=10
]{siunitx}
\usepackage{color}

\usepackage{changes}
\newcommand{\stkout}[1]{\ifmmode\text{\sout{\ensuremath{#1}}}\else\sout{#1}\fi}
\setdeletedmarkup{\stkout{#1}}

\title{{Model-Based Reconstruction for Joint Estimation of $T_{1}$,
		$R_{2}^{*}$ and $B_{0}$ Field Maps Using Single-Shot Inversion-Recovery Multi-Echo
		Radial FLASH \footnote{Part of this work has been presented at the ISMRM Workshop
			on Data Sampling and Image Reconstruction, Sedona, 2023. Xiaoqing Wang and Nick Scholand contributed equally to this work.}}}
\newcommand{\authorA}{Xiaoqing Wang}
\newcommand{\authorB}{Nick Scholand}
\newcommand{\authorC}{Zhengguo Tan}
\newcommand{\authorD}{Daniel Mackner}
\newcommand{\authorE}{Vitali Telezki}
\newcommand{\authorF}{Moritz Blumenthal}
\newcommand{\authorG}{Philip Schaten}
\newcommand{\authorH}{Martin Uecker}

\newcommand{\affilA}{Department of Radiology, Harvard Medical School, Boston, Massachusetts, USA}
\newcommand{\affilB}{Institute of Biomedical Imaging, Graz University of Technology, Austria}
\newcommand{\affilD}{German Centre for Cardiovascular Research (DZHK),
	Partner Site Göttingen, Göttingen, Germany}
\newcommand{\affilC}{Department of Artificial Intelligence in Biomedical Engineering, University of Erlangen-Nuremberg, Germany}
\newcommand{\affilE}{Institute for Diagnostic and Interventional Radiology,
	University Medical Center Göttingen, Göttingen, Germany}
\newcommand{\affilF}{Cluster of Excellence "Multiscale Bioimaging: from Molecular Machines to Networks of Excitable Cells" (MBExC), University of Goettingen, Germany}
\newcommand{\affilG}{BioTechMed-Graz, Graz, Austria}

\newcommand{\corAdress}{Xiaoqing Wang, Boston Children's Hospital, Harvard Medical School, 300 Longwood Avenue, 02115, Boston, MA, USA. }

\newcommand{\corMail}{xiaoqingwang2010@gmail.com}
\newcommand{\runHead}{Author et al.}

\iftoggle{ARXIV}{
		\author[1]{\authorA \thanks{\corAdress \corMail}}
	}

\iftoggle{MRM}{
		\author[1]{\authorA}
	}

\author[2]{\authorB}
\author[3]{\authorC}
\author[2]{\authorD}
\author[4]{\authorE}
\author[2]{\authorF}
\author[2]{\authorG}
\author[2,3,4,5,6]{\authorH}

\affil[1]{\affilA}
\affil[2]{\affilB}
\affil[3]{\affilC}
\affil[4]{\affilD}
\affil[5]{\affilE}
\affil[6]{\affilF}
\affil[7]{\affilG}
	
\iftoggle{MRM}{
	\papertype{Note}
	\corraddress{\corAdress}
	\corremail{\corMail\\
			
			Submitted to XXX
			\bf{{Word count}}: Abstract , Body .}
	
	\runningauthor{\runHead}
}

\begin{document}
	
\maketitle

\begin{abstract}
\noindent 
\textbf{Purpose}: To develop a model-based nonlinear reconstruction for simultaneous
water-specific $T_{1}$, $R_{2}^{*}$, $B_{0}$ field and/or fat fraction (FF) mapping
using single-shot inversion-recovery (IR) multi-echo radial FLASH.

\noindent \textbf{Methods}: The proposed model-based reconstruction jointly
estimates water-specific $T_{1}$, $R_{2}^{*}$, $B_{0}$ field and/or FF maps, as well as
a set of coil sensitivities directly from $k$-space obtained with a single-shot IR
multi-echo radial FLASH sequence using blip gradients across echoes. Joint sparsity
constraints are exploited on multiple quantitative maps to improve precision.
Validations are performed on numerical and NIST phantoms and with
in vivo studies of the human brain and liver at 3 T.

\noindent
\textbf{Results}: Numerical phantom studies demonstrate the effects of fat signals in $T_{1}$
estimation and confirm good quantitative accuracy of the proposed method for all
parameter maps. NIST phantom results confirm good quantitative $T_{1}$ and
$R_{2}^{*}$ accuracy in comparison to Cartesian references. Apart from good
quantitative accuracy and precision for multiple parameter maps, in vivo studies
show improved image details utilizing the proposed joint estimation. The proposed
method can achieve simultaneous water-specific $T_{1}$, $R_{2}^{*}$, $B_{0}$
field and/or FF mapping for brain (0.81 $\times$ 0.81 $\times$ 5 mm$^{3}$) and
liver (1.6 $\times$ 1.6 $\times$ 6 mm$^{3}$) imaging within four seconds.

\noindent \textbf{Conclusion}: 
The proposed model-based nonlinear reconstruction, in combination with a single-shot IR
multi-echo radial
FLASH acquisition, enables joint estimation of accurate water-specific $T_{1}$,
$R_{2}^{*}$, $B_{0}$ field and/or FF maps within four seconds. The present work is of
potential value for specific clinical applications.

\noindent
\textbf{Keywords}: model-based reconstruction, water-specified $T_{1}$ mapping, $R_{2}^{*}$ mapping, $B_{0}$ estimation, fat fraction, multi-echo radial FLASH
\end{abstract}

\section{Introduction}

Quantitative $T_{1}$ and $R_{2}^{*}$ mapping are of increasing interest 
in a number of clinical use cases \cite{Margaret_J.Magn.Reson.Imaging_2012}. 
%	For example, $T_{1}$ is useful for the diagnosis of multiple sclerosis, liver and myocardial fibrosis, and $R_{2}^{*}$ constitute a non-invasive way for quantifying iron overload. 
While $T_{1}$ relaxation time can be efficiently measured with an inversion-recovery 
(IR) Look-Locker (LL) technique \cite{Look_Rev.Sci.Instrum._1970, 
	Deichmann_J.Magn.Reson._1992, Messroghli_Magn.Reson.Med._2004},  $R_{2}^{*}$
mapping is accomplished by the multi-echo FLASH readout
\cite{Frahm_Magn.Reson.Med._1986}.  The latter can also be utilized for water-fat separation 
\cite{hernando_Magn.Reson.Med._2008, Benkert_Magn.Reson.Med._2017, armstrong_Magn.Reson.Med._2018, Schneider_Magn.Reson.Med._2020, mayer_Magn.Reson.Med._2022} based on the Dixon method \cite{Dixon_Radiology_1984}. 
Conventionally, IR LL $T_{1}$ mapping employs a single-echo readout with the shortest TE to minimize
off-resonance effects \cite{Deichmann_J.Magn.Reson._1992}. Recent advances extend the single-echo readout
to two or three echoes, which enables water-specific $T_{1}$ mapping and reduces errors caused by fat
components \cite{Feng_Magn.Reson.Med._2021, li_NMR_Biomed._2022}. Meanwhile, the latest efforts in MR  %FIXME: Context? Why Moreover? 
fingerprinting, multitasking and others empower simultaneous water-specific $T_{1}$ , $T_{2}$, $R_{2}^{*}$ mapping by extending single-echo to multi-echo readout \cite{Jaubert_Magn.Reson.Med._2020,
	hermann_Magn.Reson.Med._2021, WangN_Magn.Reson.Med._2022, 
	Cao_Magn.Reson.Med._2022,velasco2022simultaneous, Lima_Magn.Reson.Med.2022,
	Wang_NeuroImage_2022,  
	roberts2023confounder} .

% 	Recently, these two sequences are combined together in one single scan for 
% 	simultaneous multi-parameter mapping. 
% 	Selected examples include water-fat separated $T_{1}$ mapping with an inversion-prepared triple-echo  
% 	readout , or joint $T_{1}$, 
% 	water-fat separation, $R_{2}^{*}$ mapping with a longer echo train following a inversion preparation . 
% 	In line with other multi-parameter mapping techniques \cite{Ma_Nature_2013}, 
% 	the above combination has gained increasing interest as they are capable of 
% 	providing complementary quantitative information of the same tissue with a 
% 	single acquisition, reducing problems such as longer acquisition time, 
% 	inter-scan misregistration caused by separate acquisitions. 

%To obtain quantitative parameter maps, existing approaches usually 
%reconstruct contrast-weighted images first and subsequently perform a nonlinear 
%fitting/matching of the reconstructed images to the underlying physics model. 

To enable efficient quantitative imaging, sparsity and subspace-constraints along the parameter dimensions have
been developed to accelerate parameter mapping in general \cite{Doneva_Magn.Reson.Med._2010,
	petzschner_Magn.Reson.Med._2011, huang_Magn.Reson.Med.2012, velikina_Magn.Reson.Med.2013,
	Zhao_Magn.Reson.Med._2015}. Following image reconstruction, quantitative parameter maps can be obtained by
performing a nonlinear 
fitting/matching of the reconstructed images to the underlying physical model.
Among them, subspace-constrained reconstruction is one of the state-of-the-art approaches
\cite{petzschner_Magn.Reson.Med._2011, Tamir_Magn.Reson.Med._2017}. 
It approximates the MR signal evolution by a linear subspace. Because the number of subspace coefficients is usually
much smaller than that of contrast-weighted images, the subspace method is very efficient. 
%For inversion-prepared multi-echo acquisition,
%coefficients first, followed by a nonlinear fitting/matching of the 
%subspace-projected images to the underlying physics model. 
%le water-specific $T_{1}$ can be estimated by firstly reconstructing 
%ater-fat separated images and followed by pixel-wise fitting.
%To enable efficient parameter mapping, existing approaches normally utilize
%the subspace approach by
The inversion-recovery part can be effectively
represented by  only a few subspace coefficients, but
the phase accumulation in long multi-echo 
readout caused by field inhomogeneities requires a large 
number of coefficients approximating the multi-echo FLASH
signal \cite{zimmermann2017accelerated, Dong_Magn.Reson.Med._2020, Wang_Philos.Trans.R.Soc.A._2021}.
This makes the 
subspace approach less efficient in the case of combined inversion-recovery multi-echo acquisition.

Nonlinear model-based reconstruction is an alternative approach for efficient quantitative MRI %FIXME: Probable th eold papers by Fessler and Sutton should also be cited
\cite{Sutton_IEEETrans.Med.Imag._2003, Block_IEEETrans.Med.Imaging_2009, Fessler_IEEESignalProcess.Mag._2010, Zhao_IEEETrans.Med.Imag._2014, Ben-Eliezer_Magn.Reson.Med._2016, hilbert2018accelerated,
	Wang_Philos.Trans.R.Soc.A._2021, scholand2023quantitative}. It reconstructs the underlying physical parameter maps directly from
$k$-space, bypassing the intermediate steps of image reconstruction and pixelwise fitting/matching. Recently,
this kind of method was extended to efficiently reconstruct water, fat and $R_{2}^{*}$ maps from
an undersampled multi-echo FLASH acquisition \cite{Benkert_Magn.Reson.Med._2017,
	Schneider_Magn.Reson.Med._2020}. Our developments further enable an additional $B_{0}$ estimation
\cite{Tan_IEEE_TMI_2023}.  However, current model-based reconstruction can not be directly applied to the
IR multi-echo data due to the contrast changes in the course of inversion recovery, in contrary to
the steady-state conditions \cite{Benkert_Magn.Reson.Med._2017, Schneider_Magn.Reson.Med._2020, Tan_IEEE_TMI_2023}.  %FIXME: Why not? 
Existing approaches \cite{Feng_Magn.Reson.Med._2021, li_NMR_Biomed._2022} instead utilize separate steps, i.e.,
first extract the multi-echo part at longer inversion time (approximating the steady-state condition) for
water-fat 
separation, and subsequently utilize the estimated parameter maps from the first step for water-fat 
separated $T_{1}$ estimation. Although this simplifies the reconstruction, the separate use of data may 
result in sub-optimal solutions as not all available data is utilized and the correlation between % FIXME: relation instead of correlation?
quantitative maps (such as $T_{1}$ and $R_{2}^{*}$) is not fully exploited.

Inspired by the above advances, in this work, we aim to develop a fully nonlinear model-based reconstruction
method for joint water-specific $T_{1}$, $R_{2}^{*}$, $B_{0}$ and/or fat fraction (FF) mapping using the IR
multi-echo radial FLASH acquisition. In particular, we first combine a 
single-shot inversion recovery sequence with a multi-echo radial 
FLASH readout. We further incorporate blip gradients across echoes to enable an efficient $k$-space coverage. Second, after modeling the 
underlying physical signal, we formulate parameter 
estimation as a nonlinear inverse problem, i.e., to jointly 
estimate water $T_{1}$, $R_{2}^{*}$, $B_{0}$ and/or FF 
maps directly from the acquired $k$-space data. Thus, the present work is capable of exploiting all 
available information for a joint estimation. Furthermore, 
similar to \cite{Funai_IEEE_Trans.Med.Imaging_2008} and 
\cite{Wang_Magn.Reson.Med._2018}, we are able to make 
use of smooth regularization on the $B_{0}$ maps and joint 
sparsity constraints on the other parameter maps to improve the 
reconstruction.

\section{Theory}

\subsection*{Sequence Design}
\label{subsec:sequence}
The single-shot IR multi-echo radial FLASH sequence is demonstrated in the Supporting Information Figure S1.
It starts with a non-selective inversion pulse, followed by a continuous 
multi-echo radial FLASH readout. To allow for efficient $k$-space coverage, radial spokes are designed to rotate
along the echo dimension using blip gradients \cite{Tan_Magn.Reson.Med._2019}. The distribution of spokes is designed in a way that radial spokes from several TRs (e.g., 9) and all echoes are equally  distributed in one $k$-space \cite{Tan_Magn.Reson.Med._2019}, with an angle $\theta_{l,m} = 2 \pi /
(N_E \cdot N_S) \cdot [(l-1) \cdot N_E + m-1]$ for the $l$th TR and the $m$th echo. %FIXME: formula seems broken % ANSWER: Fixed now
$N_E$ and $N_S$ are the number of echoes and shots (TRs)
per $k$-space frame, respectively. Spokes acquired in consecutive $k$-space frame are  %FIXED> ??: needs a definition
rotated by a small golden-angle ($\approx 68.75 ^{\circ}$) with respect to the previous one \cite{Winkelmann_IEEETrans.Med.Imag._2007} to enable a complementary coverage of $k$-space. 

\subsection*{Signal Model}
\label{subsec:signal_equation}
Based on the signal equations of IR Look-Locker and multi-echo gradient 
echo acquisitions, the signal evolution for the above process can be described by
\begin{align}
S_{\text{TI}_{n}, \text{TE}_{m}} &=
\Big[W_{ss} - (W_{ss} + W_{0}) \cdot \exp\big(-\text{TI}_{n}\cdot R_{1,W}^{*}\big)\nonumber\\
&~+ \big(F_{ss} - (F_{ss} ~~+ F_{0}) \cdot \exp\big(-\text{TI}_{n}\cdot R_{1,F}^{*}\big)\big)\cdot z_{m}\Big]\nonumber\\
&~~~~~~~~~\cdot \exp\big(\text{TE}_{m}\cdot i2\pi f_{B_{0}}\big) \cdot \exp\big(- \text{TE}_{m} \cdot R_{2}^{*}\big)\,.
\label{eq::sigmod}
\end{align}
Where $(W_{ss},W_{0}, R_{1,W}^{*})^{T}$ are the steady-state signal, 
the equilibrium-state signal and the effective $T_{1}$ relaxation 
rate for water, and $(F_{ss},F_{0}, R_{1,F}^{*})^{T}$ represent 
the same components for fat. $z_{m}$ is the 6-peak fat 
spectrum \cite{Yu_Magn.Reson.Med._2008}, $f_{B_{0}}$ represents 
the field inhomogeneity and $R_{2}^{*}$ is the effective relaxation
rate. $\text{TI}_{n}$ and $\text{TE}_{m}$ denote the $n$th inversion 
time and $m$th echo time, respectively. The sought quantitative 
parameters in equation (\ref{eq::sigmod}) are then $x_{ p} = 
(W_{ss}, W_{0}, R_{1,W}^{*}, F_{ss}, F_{0}, R_{1,F}^{*}, f_{B_{0}}, R_{2}^{*})^{T}$. In some cases, 
such as for the National Institute of Standards and Technology (NIST) phantom \cite{stupic2021standard},
where no fat component is present, the above signal equation (\ref{eq::sigmod}) can be simplified to
\begin{align}
S_{\text{TI}_{n}, \text{TE}_{m}} &=
\Big[W_{ss} - (W_{ss} + W_{0}) \cdot \exp\big(-\text{TI}_{n}\cdot R_{1,W}^{*}\big)\Big]\nonumber \\ &~~~~~~~~~\cdot\exp\big(\text{TE}_{m}\cdot i2\pi f_{B_{0}}\big) \cdot \exp\big(- \text{TE}_{m} \cdot R_{2}^{*}\big)\,.
\label{eq::sigmod2}
\end{align}
The corresponding sought quantitative 
parameters become $x_{ p} = 
(W_{ss}, W_{0}, R_{1,W}^{*}, f_{B_{0}}, R_{2}^{*})^{T}$.

\subsection*{Model-Based Reconstruction}

Combining the above physical models with the parallel imaging equation 
\cite{Pruessmann_Magn.Reson.Med._1999, Uecker_Magn.Reson.Med._2008}, 
we construct a nonlinear forward operator $F$, which maps the unknowns 
in Equation (\ref{eq::sigmod}) and the unknown coil sensitivities $C$ to the 
acquired multi-channel data $y$ at $\text{TI}_{n}$ and $\text{TE}_{m}$, i.e., 
\begin{equation}
F: x \mapsto y = \mathcal{P} \mathcal{F}C \cdot S_{\text{TI}_{n}, \text{TE}_{m}}(x_{p})~.
\label{eq::forward}
\end{equation}
Here, $\mathcal{P}$ is the sampling pattern and $\mathcal{F}$ is 
the Fourier transform.  By setting $x_{c} = (c_{1}, \dots, c_{k}, \dots,
c_{K})^{T}$ (with $c_{k}$ the individual $k$th coil sensitivity map), 
the vector of unknowns in Equation (\ref{eq::forward}) is $x = (x_{p}, x_{c})^{T}$. 
The estimation of $x$ is then formulated as an optimization problem, i.e.,
%	$$\begin{aligned}
%	\bm x = (W_{ss}, W_{0}, R_{1,W}^{*}, F_{ss}, F_{0}, R_{1,F}^{*}, f_{B_{0}}, R_{2}^{*})^{T} = \text{argmin}_{\bm x \in D} \sum_{\text{TI}}\sum_{\text{TE}}\|P\mathcal{F}C\cdot S_{\text{TI}_{n}, , \text{TE}_{m}}(x)- Y_{\text{TI}_{n}, \text{TE}_{m}}\|_{2}^{2} + R(\bm x)
%	\end{aligned}
%	$$
\begin{equation}
\hat{x} = \text{argmin}_{ x \in D} \sum_{\text{TI}}\sum_{\text{TE}}\|\mathcal{P}\mathcal{F}C\cdot S_{\text{TI}_{n}, , \text{TE}_{m}}(x)- Y_{\text{TI}_{n}, \text{TE}_{m}}\|_{2}^{2} + R(x).\
\end{equation}
Here, D is a convex set, ensuring non-negativity of all relaxation 
rates. $R(\cdot)$ is the regularization term for both parameter 
maps and coil sensitivity maps. In particular, we use joint 
$\ell_{1}$-Wavelet sparsity constraint \cite{Wang_Magn.Reson.Med._2018}
on $(W_{ss},W_{0}, R_{1,W}^{*}, F_{ss}, F_{0}, R_{1,F}^{*}, R_{2}^{*})^{T}$  or $(W_{ss},W_{0}, R_{1,W}^{*}, R_{2}^{*})^{T}$
to exploit sparsity and correlations between maps
and Sobolev regularization on 
the $f_{B_{0}}$ map \cite{Tan_IEEE_TMI_2023} and the coil sensitivity maps \cite{Uecker_Magn.Reson.Med._2008} to enforce smoothness. 
The above optimization problem is solved by IRGNM-FISTA 
\cite{Wang_Magn.Reson.Med._2018} using the Berkeley Advanced Reconstruction Toolbox (BART) \cite{Uecker__2015}. 
More details for IRGNM-FISTA can
be found in Ref.~\cite{Wang_Magn.Reson.Med._2018}. Additionally, the derivative of the nonlinear operator 
and the adjoint of the derivative which are needed in the algorithm can be found in Appendix. %FIXME: maybe move this comment () in the Appendix %Answer: Done.

%\begin{figure}[h]
%	\centering
%	\includegraphics[width=0.45\textwidth]{Figures/test}
%	\caption{}
%	\label{Fig:test}
%\end{figure}

\section{Methods}
\subsection*{Numerical Simulations}
To evaluate the accuracy of the proposed nonlinear model-based approach, a digital phantom 
with ten circular tubes and a background was simulated. The numerical phantom models has
water $T_{1}$ ranging from 200 ms to 2000 ms for tubes, and 3000 ms for the background,
and fat $T_{1}$ ranging from 300 ms to 400 ms for tubes, and 300 ms for the background.
$R_{2}^{*}$ for the tubes is ranging from 5 $s^{-1}$ to 100 $s^{-1}$, and 5 $s^{-1}$ for 
the background, and off-resonance is ranging from $-50$ Hz to $50$ Hz for the tubes, and $-50$ Hz for the background. %FIXME quite long sentence % splitted in two. maybe we can make a table?
The fat fraction (FF) was set to be constant ($20\%$) for all tubes and the background.  
The $k$-space data was derived from the analytical Fourier representation of ellipses 
with an array of eight circular receiver coils surrounding the phantom.  %FIXME: What does "without overlap" mean? %Answer: Removed.
The simulation employed an IR multi-echo radial FLASH sequence as described above with 
a base resolution of 192 pixels covering a field of view of 128 mm, 
$\text{TR}/\text{TE}_{1}/\delta\text{TE}_{1} $ = 12.7/1.6/1.6 ms, 7 echoes, 
flip angle $6^{\circ}$, and total acquisition time of 4 s. To study the effects 
of the fat signal on the accuracy of the $T_{1}$ map, simulations with IR single-echo 
FLASH sequences, utilizing $\text{TR}/\text{TE} = 5/1.15$ ms (out-of-phase) and 
$\text{TR}/\text{TE} = 5/2.3$ ms (in-phase), were also conducted. Finally, 
complex white Gaussian noise was added to the simulated $k$-spaces.
% IMHO "moderate standard deviation" did not say so much.

\subsection*{Data Acquisition}

All MRI experiments were performed on a Magnetom Skyra 3T (Siemens Healthineers, 
Erlangen, Germany).
%	\subsection*{Parameter Choice}
%	Brain study was conducted using a 20-channel head coil. 
Validation was first performed on the $T_{2}$ spheres of a NIST phantom (model version 130)
\cite{stupic2021standard}. 
Phantom and brain studies were conducted with a
20-channel head/neck coil, whereas abdominal scans were
performed with a combined thorax and spine coil with 26
channels. Three subjects (3 male, 29 $\pm$ 7 years old)
without known illness were recruited and scans were performed after
written informed consent was obtained.
The acquisition parameters for phantom and brain were: matrix
size= 256 $\times$ 256, slice thickness = 5 mm, 7 echoes with TR = 15.6 ms,
TE$_{1-7}$ = 2.26/4.2/6.14/8.08/10.10/12.1/14.1 ms, FA = $6^{\circ}$,
bandwidth = 810 Hz/pixel, and 300 RF excitations with 2100 radial acquired
spokes for all echoes.
The FOV was 220 $\times$ 220 mm$^{2}$ for the phantom and 208 $\times$ 208 mm$^{2}$ for the brain.
Acquisition parameters for abdomen were: FOV = 320 $\times$ 320 mm$^{2}$, matrix
size= 200 $\times$ 200, slice thickness = 6 mm, 7 echoes with TR = 10.6 ms,
TE$_{1-7}$ = 1.49/2.61/3.73/4.85/5.97/7.09/8.21 ms, FA = $6^{\circ}$,
bandwidth = 1320 Hz/pixel and 360 RF excitations with 2520 radial acquired 
spokes for all echoes. Abdominal experiments were accomplished during a brief breathhold.
All acquisition parameters are summarized in the Supporting Information Table S1.

For reference, a gold-standard $T_{1}$ mapping \cite{Barral_Magn.Reson.Med._2010} was 
performed on the NIST phantom using the IR spin-echo method with 7 TIs (TI = 30, 280, 
530, 780, 1030, 1280, 1530 ms). Reference phantom and brain $R_{2}^{*}$ and $B_{0}$ 
maps were taken form the Cartesian multi-echo FLASH acquisitions provided by the vendor with TEs = (2.54, 4.92, 7.38, 
9.84, 12.30, 14.76, 17.22 ms) and  TEs = (4.92, 7.38 ms), respectively. For liver studies, reference 
$R_{2}^{*}$, fat fraction (FF) and $B_{0}$ maps were estimated using a recently 
developed model-based reconstruction of the steady-state multi-echo data \cite{Tan_IEEE_TMI_2023}. 
The steady-state data was extracted from the last 120 excitations of the same IR LL multi-echo experiment 
with a prolonged acquisition of 6 seconds and corresponding to around 1.2 s at the end. 
In-vivo $T_{1}$ references were
using the previously developed single-shot $T_{1}$ mapping method \cite{Wang_Magn.Reson.Med._2018} 
with single-echo readout.

\subsection*{Iterative Reconstruction}
The proposed model-based reconstruction
algorithm was implemented in C and CUDA using the nonlinear operator 
and optimization framework of BART \cite{blumenthal_MRM_2023}.
Both 5-parameter and 8-parameter model-based reconstructions were implemented. 
While the former was used for the NIST phantom study, the latter was 
used for reconstructing all the other data sets. Similar to \cite{Tan_IEEE_TMI_2023}, 
to prevent the phase modulation caused by phase wrapping along the echoes, 
the $B_{0}$ map was initialized from a three-point
model-based water/fat separation \cite{Tan_Magn.Reson.Med._2019} and 
the $R_{2}^{*}$ map was initialized to zero. All the other parameters 
were initialized to one. Furthermore, since $T_{1}$, fat signals, 
$R_{2}^{*}$ and $B_{0}$ have different physical units, 
they may have differently scaled gradients during optimization, 
affecting the convergence speed of the proposed method. To balance the relative gradients, 
three additional scaling variables were introduced: $L_{F}$, $L_{R_{2}^{*}}$ and $L_{B_{0}}$, 
resulting in transformed variables $\hat{F}_{ss} = F_{ss}/L_{F}$, 
$\hat{F}_{0} = F_{0}/L_{F}$, $\hat{R}_{2}^{*} = R_{2}^{*}/L_{R_{2}^{*}}$ 
and $\hat{B}_{0} = B_{0} / L_{B_{0}}$ which are used in the optimization.
Note here that both $F_{ss}$  and $F_{0}$ are scaled by $L_{F}$ according to Equation~(\ref{eq::sigmod2}). 
Practically, $L_{F}$ was set to be 0.5, $L_{R_{2}^{*}}$ and $L_{B_{0}}$ 
were set to be in the order of [0.04, 0.1] after examining the L2-norm 
of the gradient of each map during iterations manually. %FIXME: manually, automatically? % Answer: manually, added.
This setting resulted  in stable reconstructions for all cases tested, 
as shown in the convergence curves in the Supporting Information Figure S2.
%FIXME: This curves say objective function?  Is this true? Or is the this the data fidelity? $ Answer: it is data fidelity, has modified that inthe supporting information figure S2.

%partial derivatives of the cost function with respect to each parameter.

All data processing was conducted offline with BART. Following gradient delay correction using RING \cite{Rosenzweig_Magn.Reson.Med._2018a},
the acquired inversion-prepared multi-coil multi-echo data was compressed to eight virtual coils via principal component
analysis \cite{Huang_Magn.Reson.Imaging_2008}. %FIXME: citation % Answer: added
Similar to the previous studies \cite{Uecker_Magn.Reson.Med._2010, wang2023free_MRM}, 
the multicoil radial data were initially gridded onto a Cartesian grid, and all subsequent iterations 
were then performed using FFT-based convolutions with the point-spread function \cite{Wajer__2001}. %FIXME: citation Wajor % Answer: added.
To enhance computational efficiency, six spokes were combined into one $k$-space frame 
for all data sets. With the above settings, all computations were executed on a GPU 
with 48 GB of memory (RTX A6000, NVIDIA, Santa Clara, CA), resulting in a computation 
time of 10 - 20 minutes per dataset.
% Initialization
%	Initialization is very important for convergence for Newton-type algorithm 
%	such as the one employed in this work.
%	Initialization was performed 

% Scaling

% Regularization

\subsection*{Quantitative Analysis}

All quantitative results are reported as mean $\pm$ standard deviation (SD). 
Regions-of-interest (ROIs) were drawn into the quantitative maps using
% removed 'carefully'
the arrayShow \cite{Sumpf__2013} tool implemented in
MATLAB (MathWorks, Natick, MA). Bland–Altman analyses were used to compare 
ROI-based mean quantitative
values between reference and the proposed model-based reconstruction. Scan-rescan test were carried out 
to assess repeatability of this method. Moreover, the coefficient of variation (CoV = ${\text{SD}_{\text{ROI}}}/{\text{Mean}_{\text{ROI}}} \times 100\%$) was used for the evaluation of quantitative precision.%FIXME: does the later make sense?
%	Moreover, the two-tailed Student's t-tests were also utilized for quantitative comparison, and a value < 0.05 was considered significant.

\section{Results}
\subsection*{Numerical Simulation}

We first validated the proposed model-based reconstruction on a numerical phantom,  
which provides a wide range of ground-truth quantitative values in the presence of noise. 
Figure 1 (A) shows the estimated water-specific $T_{1}$, fat $T_{1}$, 
fat fraction (FF), $R_{2}^{*}$ and $B_{0}$ maps. Figure 1 (B) shows a comparison
of the corresponding ROI-analyzed quantitative values to the ground truth. The low mean differences 
($2 \pm 3$ ms, $-2 \pm 9$ ms, $0.9 \pm 1.2 $ $\%$, $-0.2 \pm 0.1$ $\text{s}^{-1}$ 
and $0.0 \pm 0.04 $ $\text{Hz}$ for water-specific $T_{1}$, fat $T_{1}$, 
fat fraction (FF), $R_{2}^{*}$ and $B_{0}$ maps, respectively) indicate
that a good quantitative accuracy has been achieved.
Figure 1 (C) shows the estimated in-phase and out-of-phase $T_{1}$ maps
using the single-echo model-based reconstruction \cite{Wang_Magn.Reson.Med._2018} 
and their comparison to the ground truth. If the fat component 
is present but not incorporated into the signal model, the mean $T_{1}$ difference
is lower than the ground truth (underestimation) in the in-phase case and higher than the ground truth (overestimation)
in the out-of-phase case. Figure 1 (D) shows the effects of the amount of fat signals
on single-echo $T_{1}$ estimation of three selected tubes. As expected, the relative
$T_{1}$ error increases with a larger fat fraction.  Note that for $T_{1}$ values
smaller than the fat $T_{1}$ (i.e., 200 ms vs. 300 ms), the difference exhibits an 
opposite sign to the larger ones.
%	 By incorporating fat signals in the reconstruction, good $T_{1}$ accuracy 
\begin{figure}[H]
	\centering
	\includegraphics[width=1.0\textwidth]{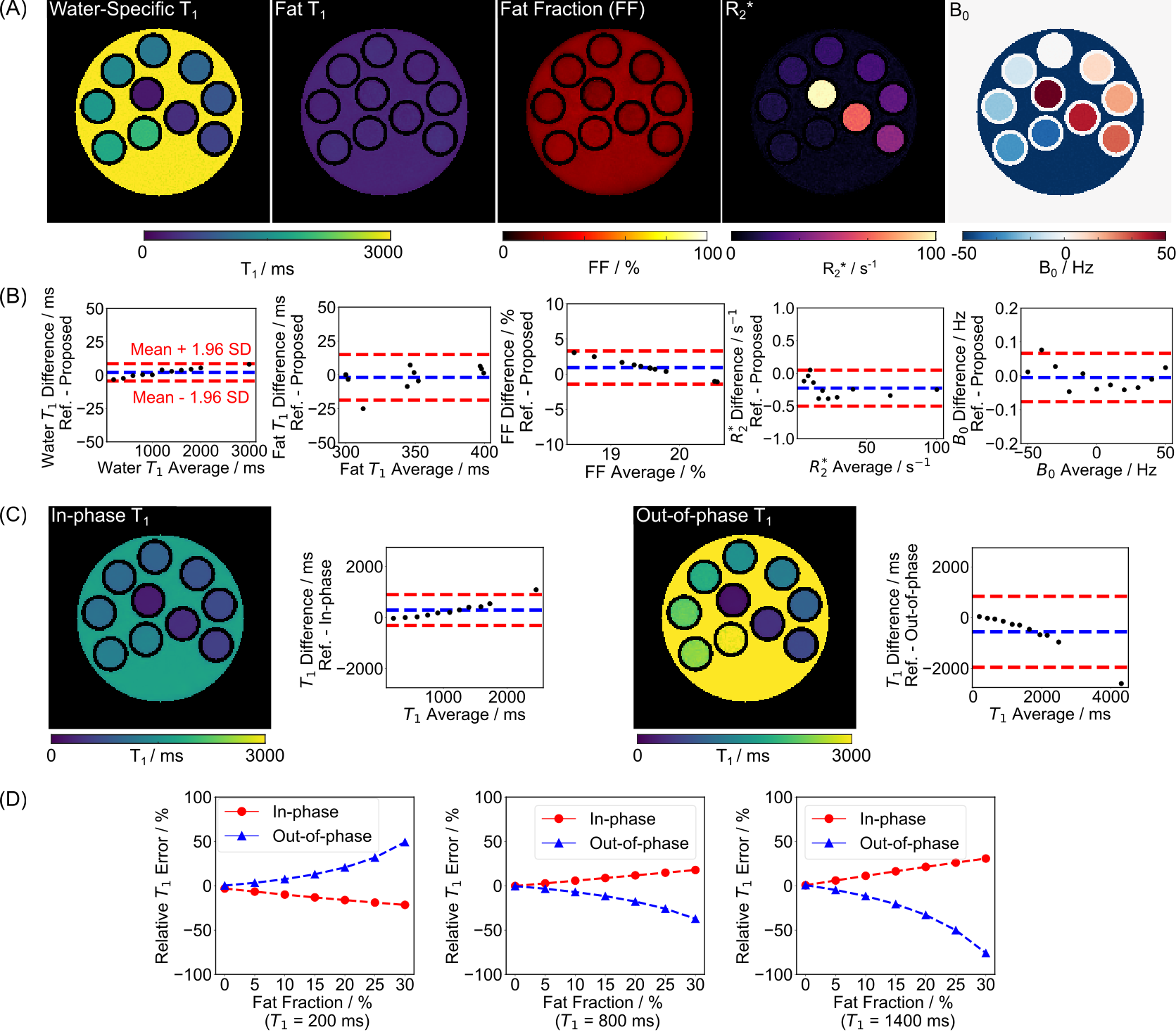}
	\caption{(A) Model-based estimated water-specific $T_{1}$, 
		fat $T_{1}$, fat fraction (FF), $R_{2}^{*}$ and $B_{0}$ maps from
		a single-shot multi-echo radial acquisition of a numerical phantom. 
		(B) Bland–Altman plots comparing the ROI-analyzed mean quantitative 
		values to the ground truth. The mean differences are $2 \pm 3$ ms, 
		$-2 \pm 9$ ms, $0.9 \pm 1.2 $ $\%$, $-0.2 \pm 0.1$ $\text{s}^{-1}$ 
		and $0.0 \pm 0.04 $ $\text{Hz}$ for water-specific $T_{1}$, fat $T_{1}$, 
		FF, $R_{2}^{*}$ and $B_{0}$ maps, respectively. (C) In-phase 
		($\text{TE} = 2.3$ ms) and out-of-phase ($\text{TE} = 1.15$ ms)
		$T_{1}$ maps estimated using the single-echo model-based reconstruction
		\cite{Wang_Magn.Reson.Med._2018}. The mean differences are 
		$292 \pm 309$ ms and $-558 \pm 713$ ms for in-phase and 
		out-of-phase $T_{1}$s, respectively. Similar to (A) and (B),
		the fat fraction (FF) is $20\%$ here. (D) The effects of fat signals on single-echo $T_{1}$ estimation of three selected tubes. The more fat components, the larger the relative $T_{1}$ errors. Note that for  $T_{1}$ smaller than the 
		fat $T_{1}$ (i.e., 200 ms vs. 300 ms), the difference 
		exhibits an opposite sign to the larger ones. }
\end{figure}

\subsection*{Phantom Validation}

Figure 2 (A) shows the estimated NIST phantom $T_{1}$, $R_{2}^{*}$ and $B_{0}$ maps 
using the proposed method and reference methods. Note that the 5-parameter model 
was used here. Visual inspection reveals good agreement between the proposed method 
and references for the selected tubes ($T_{1}$ range of [200, 3000] ms and $R_{2}^{*}$ 
range of [10, 200] $s^{-1}$ (equivalent to $T_{2}^{*}$ range of [5, 100] ms)). 
Moreover, the proposed method reduces background noise in the $R_{2}^{*}$ map 
due to the direct regularization on the $R_{2}^{*}$ map. The good correspondence 
is confirmed quantitatively in the ROI-analyzed values that are shown in Figure 2 (B)
with relatively small mean differences of $31 \pm 21$ ms and $0.3 \pm 1.4$ $\text{s}^{-1}$ 
for $T_{1}$ and $R_{2}^{*}$, respectively.   
\begin{figure}[H]
	\centering
	\includegraphics[width=1.0\textwidth]{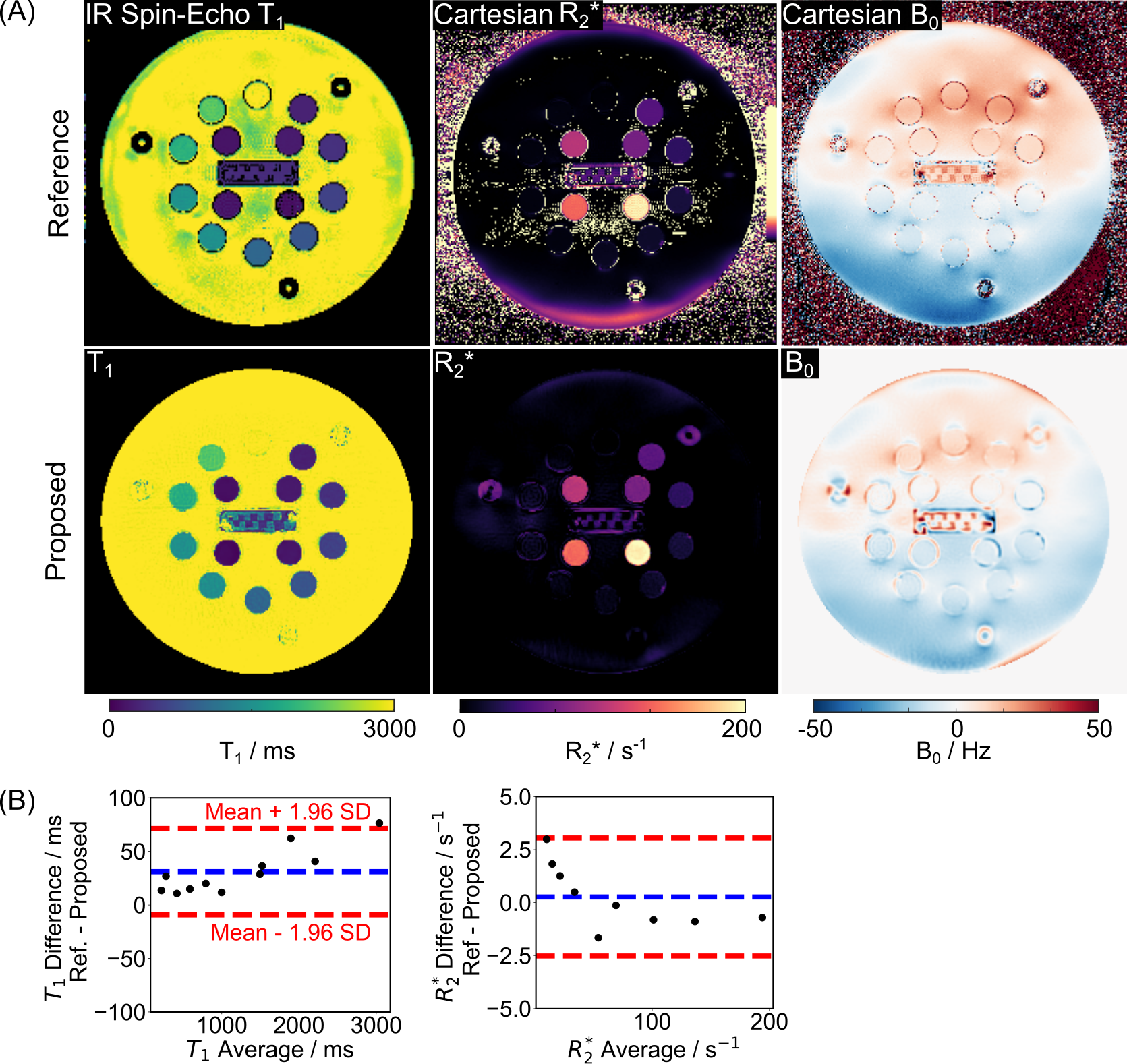}
	\caption{ (A) Model-based estimated $T_{1}$, 
		$R_{2}^{*}$ and $B_{0}$ maps and their comparison to the 
		Cartesian references of the NIST phantom. (B) Bland–Altman 
		plots comparing the ROI-analyzed mean quantitative $T_{1}$ 
		and $R_{2}^{*}$ values to the references. The mean differences 
		are $31 \pm 21$ ms and $0.3 \pm 1.4$ $\text{s}^{-1}$ for 
		$T_{1}$ and $R_{2}^{*}$ maps, respectively. Please note 
		that only tubes with a $T_{1}$ range of [200, 3000] ms 
		and $R_{2}^{*}$ range of [10, 200] $s^{-1}$ (equivalent 
		to $T_{2}^{*}$ range of [5, 100] ms) are displayed. }
\end{figure}

\subsection*{In Vivo Studies}

The Supporting Information Figure S3 (A) demonstrates the effect of the regularization parameter on brain water-specific $T_{1}$ and $R_{2}^{*}$ maps. Low values of the regularization parameter result in increased noise, while high values lead to improved precision (CoV values from left to right: $4.3\pm 0.5 \%$, $3.5\pm 0.5 \%$, $3.1\pm 0.7 \%$ and $2.8\pm 0.9 \%$ for $T_{1}$ and $10.0\pm 2.3 \%$, $6.8\pm 1.3 \%$, $5.2\pm 1.0 \%$ and $4.3\pm 1.0 \%$ for $R_{2}^{*}$), but it also causes blurring. A value of 0.004 was chosen to balance noise reduction and preservation of image details for the brain reconstruction. Similarly, the regularization parameter was set to be 0.007 for the abdominal reconstructions, as shown in the Supporting Information Figure S3 (B).

With the above settings, Figure 3 (A) compares model-based reconstructed brain quantitative parameter
maps to the references. 
Apart from variations in fluid and fat regions, the estimated quantitative 
maps are visually comparable to the references. The magnified view of 
the $T_{1}$ maps highlights improved image details (black arrow) through the
proposed joint estimation and demonstrates the removal of fat signal from 
the water-specific $T_{1}$ map (white arrow). Supporting Information 
Figure S4 (A) illustrates the corresponding jointly estimated brain 
water ($W_{ss}$) and fat ($F_{ss}$) images.
Furthermore, the proposed method exhibits fewer distortion effects on 
the $R_{2}^{*}$ map compared to the reference  method (black arrow). 
The Bland-Altman plot of ROI-analyzed values in Figure 3 (B) reveals 
good agreement between the proposed method and the references. 
The mean $R_{2}^{*}$ difference is $-0.1 \pm 1.4$ $\text{s}^{-1}$, 
and the mean $T_{1}$ difference is  $-16 \pm 8$ ms. The latter 
indicates slightly higher water-specific $T_{1}$ values by the 
proposed method compared to those estimated from the single-shot
IR single-echo acquisition.
\begin{figure}[H]
	\centering
	\includegraphics[width=1.0\textwidth]{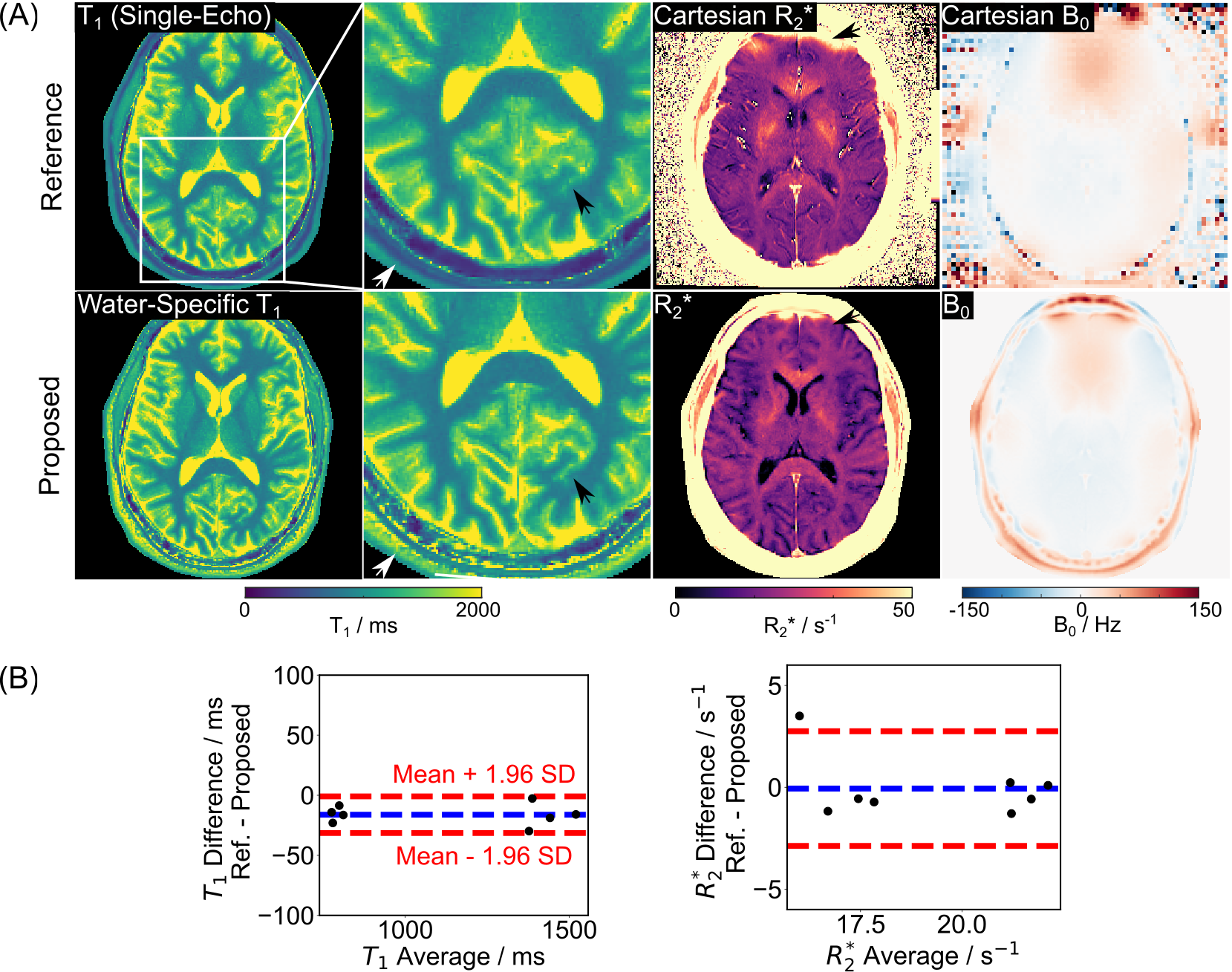}
	\caption{(A) Model-based reconstructed brain 
		water-specific $T_{1}$, its enlarged region, $R_{2}^{*}$ 
		and $B_{0}$ maps and their comparison to the references. 
		The white arrow on the $T_{1}$ map indicates fat signal 
		removal, while the black arrow highlights better preserved 
		structural details with joint estimation. The black arrow 
		on the $R_{2}^{*}$ maps reflects reduced distortion artifacts
		achieved by the proposed method. (B) Bland–Altman plots 
		comparing the ROI-analyzed mean quantitative water-specific 
		$T_{1}$ and $R_{2}^{*}$ values to the references. The mean 
		differences are $-16 \pm 8$ ms and $-0.1 \pm 1.4$ $\text{s}^{-1}$ 
		for water-specific $T_{1}$ and $R_{2}^{*}$ maps, respectively.}
\end{figure}

%	Moreover, the water steady-state ($W_{ss}$ in equation ) and fat steady-state maps shown in supporting information figure indicates a good 

%	For the sake of completeness, the estimated water steady-state ($W_{ss}$ in equation ) and fat steady-state maps are presented in the supporting figure S2. 

Figure 4 (A) displays two repetitive brain water-specific  $T_{1}$, 
$R_{2}^{*}$, and $B_{0}$ maps generated using the proposed method 
for two subjects. The minor visual differences between repetitive 
scans demonstrate good intra-subject repeatability. Quantitative 
confirmation in Figure 4 (B) via the Bland-Altman plot reveals 
small mean differences of $4 \pm 11$ ms and $0.1 \pm 0.5$ $\text{s}^{-1}$ 
for water-specific $T_{1}$ and $R_{2}^{*}$, respectively.
\begin{figure}[H]
	\centering
	\includegraphics[width=1.0\textwidth]{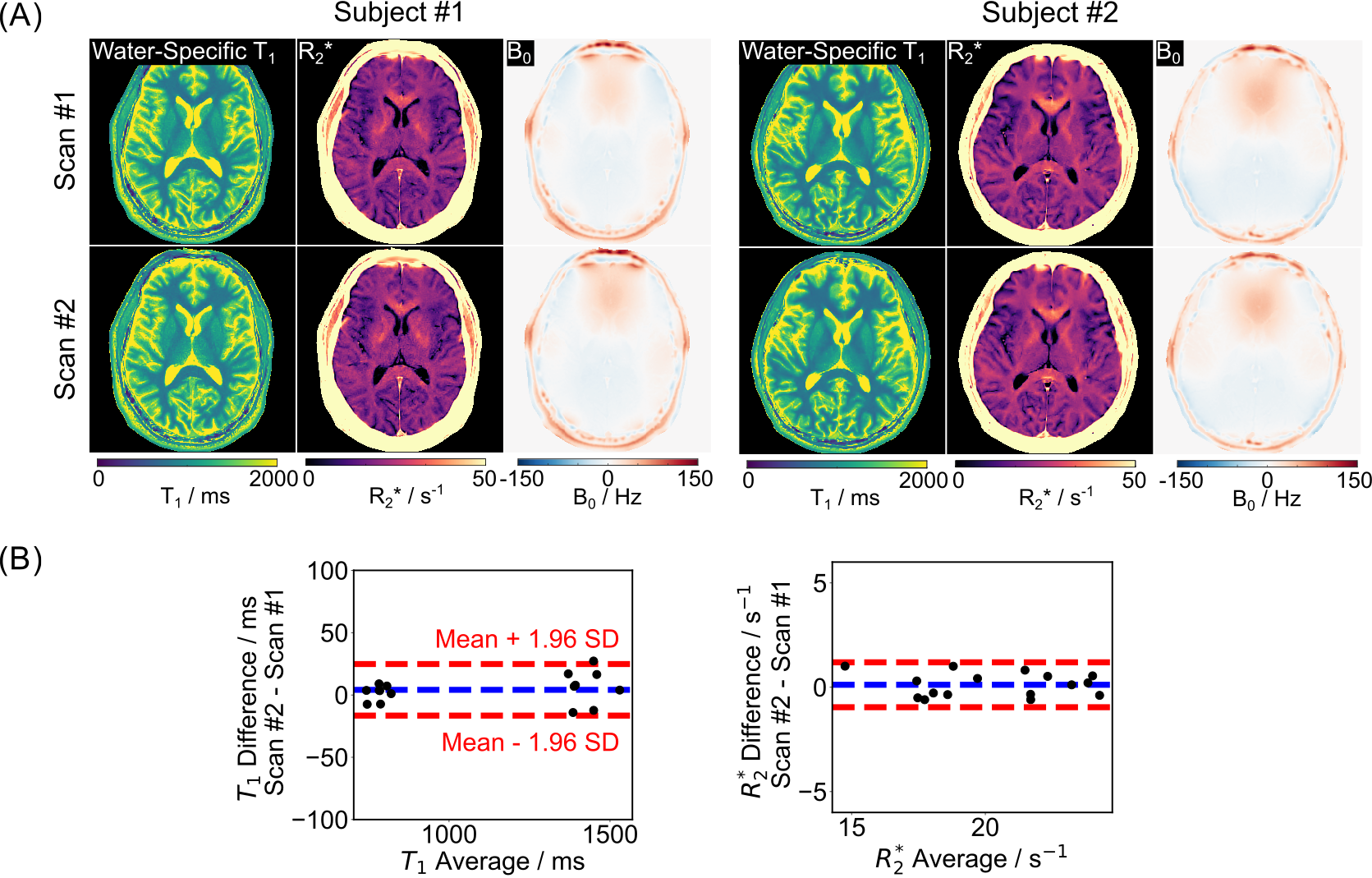}
	\caption{(A)Model-based reconstructed brain water-specific $T_{1}$, $R_{2}^{*}$ and $B_{0}$ maps for two repetitive scans and two subjects. (B) Bland–Altman plots comparing the ROI-analyzed mean quantitative water-specific $T_{1}$ and $R_{2}^{*}$ values for two repetitive scans and two subjects. The mean differences are $4 \pm 11$ ms and $0.1 \pm 0.5$ $\text{s}^{-1}$ for $T_{1}$ and $R_{2}^{*}$ maps, respectively.}
\end{figure}
%	Figure 5 (A) shows model-based reconstructed liver water-specific $T_{1}$, $R_{2}^{*}$, fat fraction and $B_{0}$ maps using the proposed method and reference methods for one representative subject. Although the breathing conditions are slightly different between scans, quantitative
%	maps are visually comparable between the proposed method and the reference methods. Similar to the brain case, the proposed joint estimation not only improves the spatial delineation of image details, especially for the $R_{2}^{*}$ map (white arrows), but also removes the fat signal in the water-specific $T_{1}$ map (white arrow) (the jointly estimated liver water $W_{ss}$ and fat $F_{ss}$ images are shown in the supporting information figure 4). Despite of the above differences, the quantitative comparison in Figure 5 (B) shows good agreement with mean $R_{2}^{*}$  difference of $-0.8 \pm 4$ $\text{s}^{-1}$, FF difference of $-0.2 \pm 0.9$ $\%$ and mean $T_{1}$ difference of $13 \pm 22$ ms. Again, the latter reflects that the water-specific $T_{1}$ values by the proposed method are slightly different (higher in this case) from the one estimated from the single-shot IR single-echo acquisition. In line with the brain results, the quantitative comparison in Figure 5 (C) shows small mean differences between two repetitive scans with mean difference of $-0.5 \pm 6$ ms, $-2.4 \pm 3.3$ $\text{s}^{-1}$, and $0.3 \pm 0.8$ $\%$ for water-specific $T_{1}$ and $R_{2}^{*}$, and fat fraction, indicating good repeatability of the proposed method.

Figure 5 (A) presents model-based reconstructed liver water-specific $T_{1}$, $R_{2}^{*}$, 
fat fraction, and $B_{0}$ maps using the proposed method and reference methods for 
one subject.  Despite slight differences in breathing conditions 
between the two scans, quantitative maps are comparable. Similar to the brain case, 
the proposed joint estimation enhances spatial delineation of image details, 
particularly in the $R_{2}^{*}$ map (white arrows), and eliminates fat signal 
in the water-specific  $T_{1}$ map (white arrow). Supporting Information 
Figure S4 (B) includes the corresponding jointly estimated liver 
water ($W_{ss}$) and fat ($F_{ss}$) images.
The quantitative comparison in Figure 5 (B) reveals good agreement  
with mean differences of $-0.6 \pm 3$ $\text{s}^{-1}$ for $R_{2}^{*}$, 
$-0.3 \pm 1.0$ \% for fat fraction (FF), and $12 \pm 21$ ms for water-specific $T_{1}$. 
Again, the latter reflects slight differences (higher in this case) in 
water-specific $T_{1}$ values by the proposed method compared to those 
estimated from the single-shot IR single-echo acquisition. Additionally, the Supporting
Information Figure S5 shows the reconstructed quantitative maps for the second scan. 
Despite differences in the $B_{0}$ maps to the first scan, which are likely caused
by breathing \cite{Tan_Magn.Reson.Med._2019}, all other quantitative maps show good visual agreement. 
The quantitative comparison in Figure 5 (C) demonstrates 
small mean differences, 
with mean differences of $0.6 \pm 6$ ms, $-2.5 \pm 3.5$ $\text{s}^{-1}$, 
and $0.2 \pm 0.8$ $\%$ for water-specific $T_{1}$, $R_{2}^{*}$, and fat fraction, 
respectively, indicating good repeatability of the proposed method. 
\begin{figure}[H]
	\centering
	\includegraphics[width=1.0\textwidth]{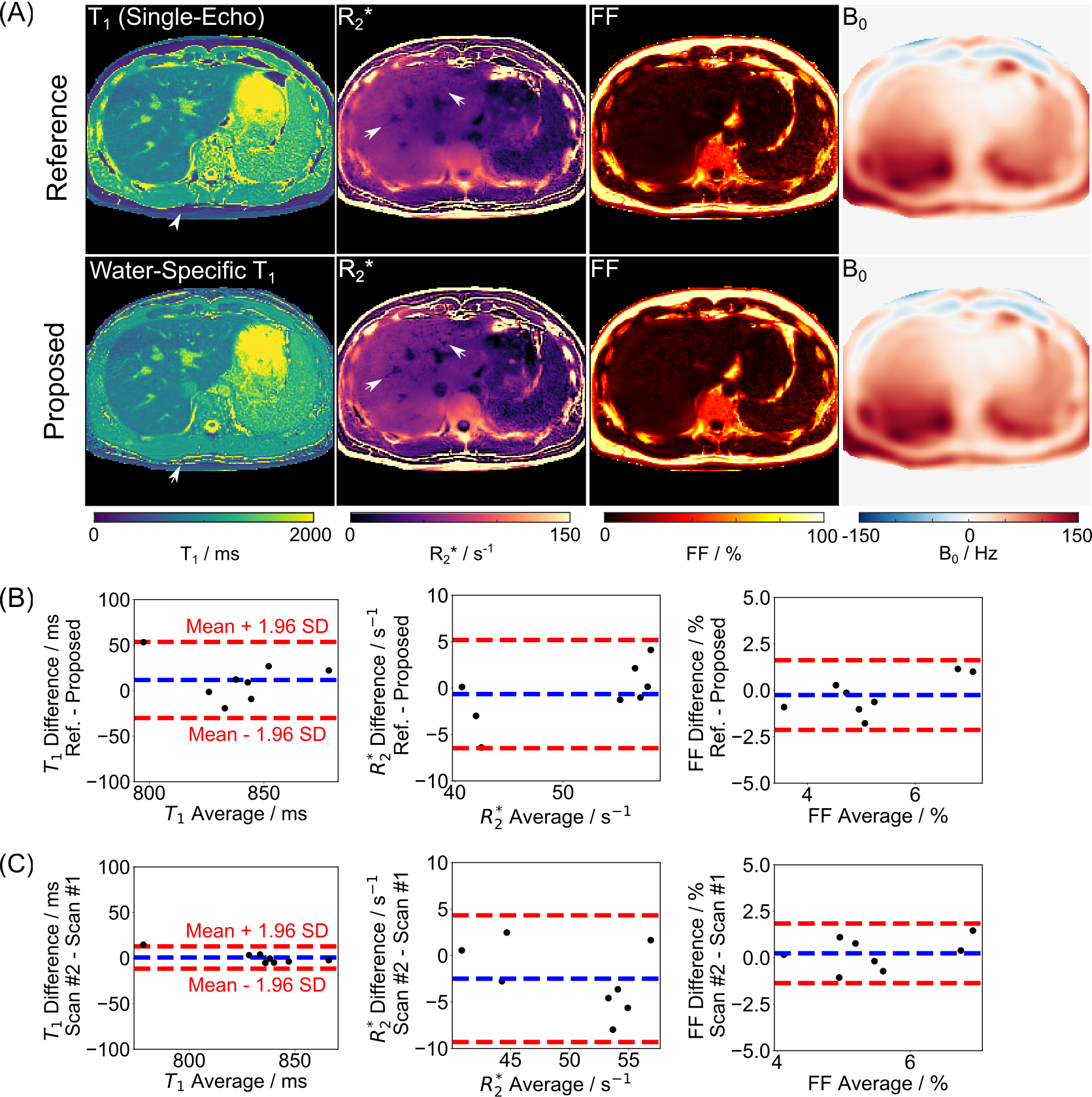}
	\caption{(A) Model-based reconstructed liver water-specific 
		$T_{1}$, $R_{2}^{*}$, fat fraction and $B_{0}$ maps and their comparison
		to the references. The white arrow on the $T_{1}$ map indicates the 
		removal of fat signal, while the white arrows on the $R_{2}^{*}$ maps
		reflect improved structure details achieved with joint estimation.
		(B) Bland–Altman plots comparing the ROI-analyzed mean quantitative
		water-specific $T_{1}$, $R_{2}^{*}$ and fat fraction values to 
		the references. The mean differences are $12 \pm 21$ ms, 
		$-0.6 \pm 3$ $\text{s}^{-1}$ and $-0.3 \pm 1.0$ $\%$  for 
		water-specific $T_{1}$,  $R_{2}^{*}$ and FF maps, respectively.
		(C) Bland–Altman plots comparing ROI-analyzed mean quantitative
		values for two repetitive scans of the same subject. The mean
		differences are $0.6 \pm 6$ ms, $-2.5 \pm 3.5$ $\text{s}^{-1}$
		and $0.2 \pm 0.8 \%$ for water-specific $T_{1}$, $R_{2}^{*}$ and
		FF maps, respectively. Quantitative maps for scan $\#$2 can be 
		found in the Supporting Information Figure S5.  }
\end{figure}

\section{Discussion}
	In this work, we investigated a joint water-specific $T_{1}$, $R_{2}^{*}$, $B_{0}$ and
FF mapping technique using single-shot IR multi-echo radial FLASH 
acquisition and nonlinear calibrationless model-based reconstruction.
%	In this work, we have developed a nonlinear calibrationless model-based reconstruction for joint estimation of water-specific $T_{1}$, $R_{2}^{*}$, fat fraction, $B_{0}$ filed mapping technique using a single-shot IR radial multi-echo FLASH acquisition. 
%	With complementary radial spokes covering $k$-space across echoes,
The proposed method models the underlying physical MR signal and formulates parameter
estimation directly from all acquired data as a nonlinear inverse problem. 
This formulation allows the use of all complementary $k$-space information to
determine the underlying quantitative maps. It also enables the use
of joint sparsity constraints for multiple quantitative maps to further improve 
precision. Validation studies on a numerical phantom, the NIST phantom, 
brain and abdominal studies have not only demonstrated good quantitative
accuracy, precision and repeatability, but also illustrated improved image
details.

The enhanced image details may mainly result from leveraging all available information obtained through the IR multi-echo acquisition. 	
In our current brain studies, for instance, utilizing 300 RF excitations and 
7 echoes, a total of 2100 distinct spokes are exploited for the joint parameter estimation. This total number of 
spokes is more than that of the IR single-echo case \cite{Wang_Magn.Reson.Med._2018},
using around 975 spokes (with $\text{TR} = 4.1$ ms) for $T_{1}$ estimation. 
For the abdominal case, while the steady state has 840 ($120 \times 7$) 
distinct spokes, the proposed method uses 2520 ($360 \times 7$) different 
spokes from the entire inversion recovery for joint estimation. 

The present work is situated within the ongoing investigation of 
rapid confounder-corrected $T_{1}$ mapping and joint multi-parameter 
mapping \cite{Feng_Magn.Reson.Med._2021, li_NMR_Biomed._2022, Jaubert_Magn.Reson.Med._2020,
	hermann_Magn.Reson.Med._2021, WangN_Magn.Reson.Med._2022,velasco2022simultaneous,Cao_Magn.Reson.Med._2022,Lima_Magn.Reson.Med.2022, roberts2023confounder, zimmermann2023qrage, heydari2024joint}.  It has been reported
that $T_{1}$ accuracy is affected by the presence of fat 
\cite{mozes2016influence, Feng_Magn.Reson.Med._2021, li_NMR_Biomed._2022,WangN_Magn.Reson.Med._2022,roberts2023confounder}. 
In line with previous findings, our simulation results confirm that,
if fat is present but not modeled, $T_{1}$ values that are larger than fat
$T_{1}$ are underestimated in the in-phase case and overestimated 
in the out-of-phase case. The relative $T_{1}$ estimation error increases
with a higher fat fraction. Moreover, our simulation shows 
vice-versa estimation results for $T_{1}$ that are smaller than 
the fat $T_{1}$. However, when the fat component is incorporated 
in the signal model, accurate $T_{1}$ can be determined using the 
proposed joint estimation. Our model-based brain and liver results 
have shown a similar trend. I.e., the water-specific brain $T_{1}$ 
is slightly higher than the single-echo case ($\text{TE}$ = 2.36 ms,
close to the in-phase condition) and liver $T_{1}$ is slightly 
lower than the single-echo case ($\text{TE}$ = 1.49 ms, close to the out-of-phase condition).

As a technical development, the main limitation of the current study 
is the small sample size, and that no patient studies have been conducted.
Evaluating the approach on a larger cohort, including both volunteers 
and patients, especially those with fatty liver disease and/or iron 
overload, would be of great interest and will be the subject of future work.
The proposed model-based reconstruction is very general and can be
applied to 3D imaging as well. Although the focus of the study has been on 2D
imaging with a efficient single-shot acquisition, future research, 
integrating the proposed model-based reconstruction with techniques 
such as simultaneous multi-slice excitation \cite{Rosenzweig_Magn.Reson.Med._2018}
or stack-of-stars acquisition \cite{Feng_Magn.Reson.Med._2021, Tan_IEEE_TMI_2023}, 
holds significant promise as it could enable efficient multiple parameter mapping 
of the entire brain and/or liver.  %FIXME: probably needs to mention other literature here in put it into context.

\section{Conclusion}
	The proposed model-based nonlinear reconstruction, in combination of 
a single-shot IR multi-echo radial FLASH acquisition, enables a joint 
estimation of accurate water-specific $T_{1}$, $R_{2}^{*}$, $B_{0}$,
and/or FF maps with good precision and repeatability. The present work 
is of potential value for specific clinical applications.

\section{Open Research}
\subsection{Data Availability Statement}
In the spirit of reproducible research, code to
reproduce the reconstruction and analysis will be available on
\url{https://github.com/mrirecon/moba-irme}.
The raw $k$-space data, all ROIs to reproduce the quantitative values and other relevant files used in this study can be downloaded from
\url{https://doi.org/10.5281/zenodo.10529421}.
%.
\section{Acknowledgements}
We thank Dr. Li Feng from New York University for the inspiring discussions. 
\appendix
\section{Appendix}
	\subsection*{ Derivative and Adjoint Derivatives of the Operators}
%	\subsubsection{Eight-parameter model}
\begin{align*}
DF(x)\begin{pmatrix}
dW_{ss} \\
dW_{0} \\
dR_{1,w}^{*} \\
dF_{ss} \\
dF_{0} \\
dR_{1,F}^{*} \\
df_{B_{0}} \\
dR_{2}^{*} \\
dc_{1} \\
\vdots \\
dc_{K} 
\end{pmatrix} &=\left( \begin{array}{c}
\mathcal{P}_{{1}, {1}} \mathcal{F}\Big\{ c_{1} \cdot  \Big(\frac{\partial{S}_{{1}, {1}}}{\partial{W_{ss}}}dW_{ss} + \frac{\partial{S}_{{1}, {1}}}{\partial{W_{0}}}dW_{0} +  \frac{\partial{S}_{{1}, {1}}}{\partial{R_{1,w}^{*}}}dR_{1,w}^{*} +
\frac{\partial{S}_{{1}, {1}}}{\partial{F_{ss}}}dF_{ss} + 
\frac{\partial{S}_{{1}, {1}}}{\partial{F_{0}}}dF_{0} + \\
\frac{\partial{S}_{{1}, {1}}}{\partial{R_{1,F}}}dR_{1,F}^{*} +
\frac{\partial{S}_{{1}, {1}}}{\partial{f_{B_{0}}}}df_{B_{0}} +
\frac{\partial{S}_{{1}, {1}}}{\partial{R_{2}^{*}}}dR_{2}^{*}\Big) + S_{1,1} \cdot dc_{1} \Big\}\\
\vdots \\
\mathcal{P}_{{n}, {m}} \mathcal{F}\Big\{ c_{1} \cdot  \Big(\frac{\partial{S}_{{n}, {m}}}{\partial{W_{ss}}}dW_{ss} + \frac{\partial{S}_{{n}, {m}}}{\partial{W_{0}}}dW_{0} +  \frac{\partial{S}_{{n}, {m}}}{\partial{R_{1,w}^{*}}}dR_{1,w}^{*} +
\frac{\partial{S}_{{1}, {1}}}{\partial{F_{ss}}}dF_{ss} + 
\frac{\partial{S}_{{1}, {1}}}{\partial{F_{0}}}dF_{0} + \\
\frac{\partial{S}_{{1}, {1}}}{\partial{R_{1,F}}}dR_{1,F}^{*} +
\frac{\partial{S}_{{n}, {m}}}{\partial{f_{B_{0}}}}df_{B_{0}} +
\frac{\partial{S}_{{n}, {m}}}{\partial{R_{2}^{*}}}dR_{2}^{*}\Big) + S_{n,m} \cdot dc_{1} \Big\} \\
\vdots \\
\mathcal{P}_{{n}, {m}} \mathcal{F}\Big\{ c_{K} \cdot  \Big(\frac{\partial{S}_{{n}, {m}}}{\partial{W_{ss}}}dW_{ss} + \frac{\partial{S}_{{n}, {m}}}{\partial{W_{0}}}dW_{0} +  \frac{\partial{S}_{{n}, {m}}}{\partial{R_{1,w}^{*}}}dR_{1,w}^{*} +
\frac{\partial{S}_{{1}, {1}}}{\partial{F_{ss}}}dF_{ss} + 
\frac{\partial{S}_{{1}, {1}}}{\partial{F_{0}}}dF_{0} + \\
\frac{\partial{S}_{{1}, {1}}}{\partial{R_{1,F}}}dR_{1,F}^{*} +
\frac{\partial{S}_{{n}, {m}}}{\partial{f_{B_{0}}}}df_{B_{0}} +
\frac{\partial{S}_{{n}, {m}}}{\partial{R_{2}^{*}}}dR_{2}^{*}\Big) + S_{n,m} \cdot dc_{K} \Big\} \\
\vdots \\
\mathcal{P}_{{N}, {M}} \mathcal{F}\Big\{ c_{K} \cdot  \Big(\frac{\partial{S}_{{N}, {M}}}{\partial{W_{ss}}}dW_{ss} + \frac{\partial{S}_{{N}, {M}}}{\partial{W_{0}}}dW_{0} +  \frac{\partial{S}_{{N}, {M}}}{\partial{R_{1,w}^{*}}}dR_{1,w}^{*} +
\frac{\partial{S}_{{N}, {M}}}{\partial{F_{ss}}}dF_{ss} + 
\frac{\partial{S}_{{N}, {M}}}{\partial{F_{0}}}dF_{0} + \\
\frac{\partial{S}_{{1}, {1}}}{\partial{R_{1,F}}}dR_{1,F}^{*} +
\frac{\partial{S}_{{N}, {M}}}{\partial{f_{B_{0}}}}df_{B_{0}} +
\frac{\partial{S}_{{N}, {M}}}{\partial{R_{2}^{*}}}dR_{2}^{*}\Big) + S_{N,M} \cdot dc_{K} \Big\} 
\end{array} 
\right)
\label{eq:der8para}
\end{align*}

and the adjoint is

\begin{align*}
\begin{pmatrix}
dW_{ss} \\
dW_{0} \\
dR_{1,w}^{*} \\
dF_{ss} \\
dF_{0} \\
dF_{1,w}^{*} \\
df_{B_{0}} \\
dR_{2}^{*} \\
dc_{1} \\
\vdots \\
dc_{K} 
\end{pmatrix} &= DF^{H}(x)\begin{pmatrix}
y_{1, 1, 1} \\
\vdots \\
y_{K, 1,1} \\
y_{1, 2, 1}\\
\vdots \\
y_{K, N, M}
\end{pmatrix} &=\left( \begin{array}{c}
\displaystyle \sum_{k=1}^{K} \sum_{n=1}^{N}\sum_{m=1}^{M} \left(\overline{\frac{\partial{S}_{{n},{m}}}{\partial{W_{ss}}}}
\cdot \overline{c_{k}} \cdot \mathcal{F}^{-1}\{\mathcal{P}_{{n}, {m}}y_{k,n,m}\} \right) \\
\vdots \\
\displaystyle \sum_{k=1}^{K} \sum_{n=1}^{N}\sum_{m=1}^{M} \left(\overline{\frac{\partial{S}_{{n},{m}}}{\partial{F_{ss}}}}
\cdot \overline{c_{k}} \cdot \mathcal{F}^{-1}\{\mathcal{P}_{{n}, {m}}y_{k,n,m}\} \right) \\
\vdots \\
\displaystyle \sum_{k=1}^{K} \sum_{n=1}^{N}\sum_{m=1}^{M} \left(\overline{\frac{\partial{S}_{{n},{m}}}{\partial{R_{2}^{*}}}}
\cdot \overline{c_{k}} \cdot \mathcal{F}^{-1}\{\mathcal{P}_{{n}, {m}}y_{k,n,m}\} \right) \\
\displaystyle \sum_{n=1}^{N}\sum_{m=1}^{M} \left(\overline{{S}_{{n},{m}}} \cdot \mathcal{F}^{-1}\{\mathcal{P}_{{n}, {m}}y_{1,n,m}\} \right) \\
\vdots \\
\displaystyle \sum_{n=1}^{N}\sum_{m=1}^{M} \left(\overline{{S}_{{n},{m}}} \cdot \mathcal{F}^{-1}\{\mathcal{P}_{{n}, {m}}y_{K,n,m}\} \right)
\end{array} 
\right)
\end{align*}
with partial derivatives
\begin{align*}
\frac{\partial{S}_{{n},{m}}}{\partial{W_{ss}}} & = \big(1-\exp\big(-\text{TI}_{n}\cdot R_{1,W}^{*}\big)\big) \cdot Z\\
\frac{\partial{S}_{{n},{m}}}{\partial{W_{0}}} & = -\exp\big(-\text{TI}_{n}\cdot R_{1,W}^{*}\big) \cdot Z \\
\frac{\partial{S}_{{n},{m}}}{\partial{R_{1,w}^{*}}} & = \text{TI}_{n}\cdot (W_{ss} + W_{0}) \cdot \exp\big(-\text{TI}_{n}\cdot R_{1,W}^{*}\big) \cdot Z\\
\frac{\partial{S}_{{n},{m}}}{\partial{F_{ss}}} & = \big(1-\exp\big(-\text{TI}_{n}\cdot R_{1,F}^{*}\big)\big) \cdot z_{m}\cdot Z\\
\frac{\partial{S}_{{n},{m}}}{\partial{F_{0}}} & = -\exp\big(-\text{TI}_{n}\cdot R_{1,F}^{*}\big) \cdot z_{m}\cdot Z\\
\frac{\partial{S}_{{n},{m}}}{\partial{R_{1,F}^{*}}} & = \text{TI}_{n}\cdot (F_{ss} + F_{0}) \cdot \exp\big(-\text{TI}_{n}\cdot R_{1,F}^{*}\big)\cdot z_{m}  \cdot Z\\
\frac{\partial{S}_{{n},{m}}}{\partial{f_{B_{0}}}} & = i2\pi\cdot \text{TE}_{m}\cdot  S_{n,m} \\
\frac{\partial{S}_{{n},{m}}}{\partial{R_{2}^{*}}} & = -\text{TE}_{m} \cdot S_{n,m} \\
\end{align*}
where $Z = \exp\big(\text{TE}_{m}\cdot i2\pi f_{B_{0}}\big) \cdot \exp\big(-\text{TE}_{m} \cdot R_{2}^{*}\big) $, $S_{n,m}$ is the signal model in equation (\ref{eq::sigmod}) and $\bar{ }$ denotes pointwise complex conjugation. The 5-parameter ones 
can be derived by removing the fat parts from the above equations.
\bibliography{radiology}

%\iftoggle{SUPMATERIAL}
%{
\includepdf[pages=-]{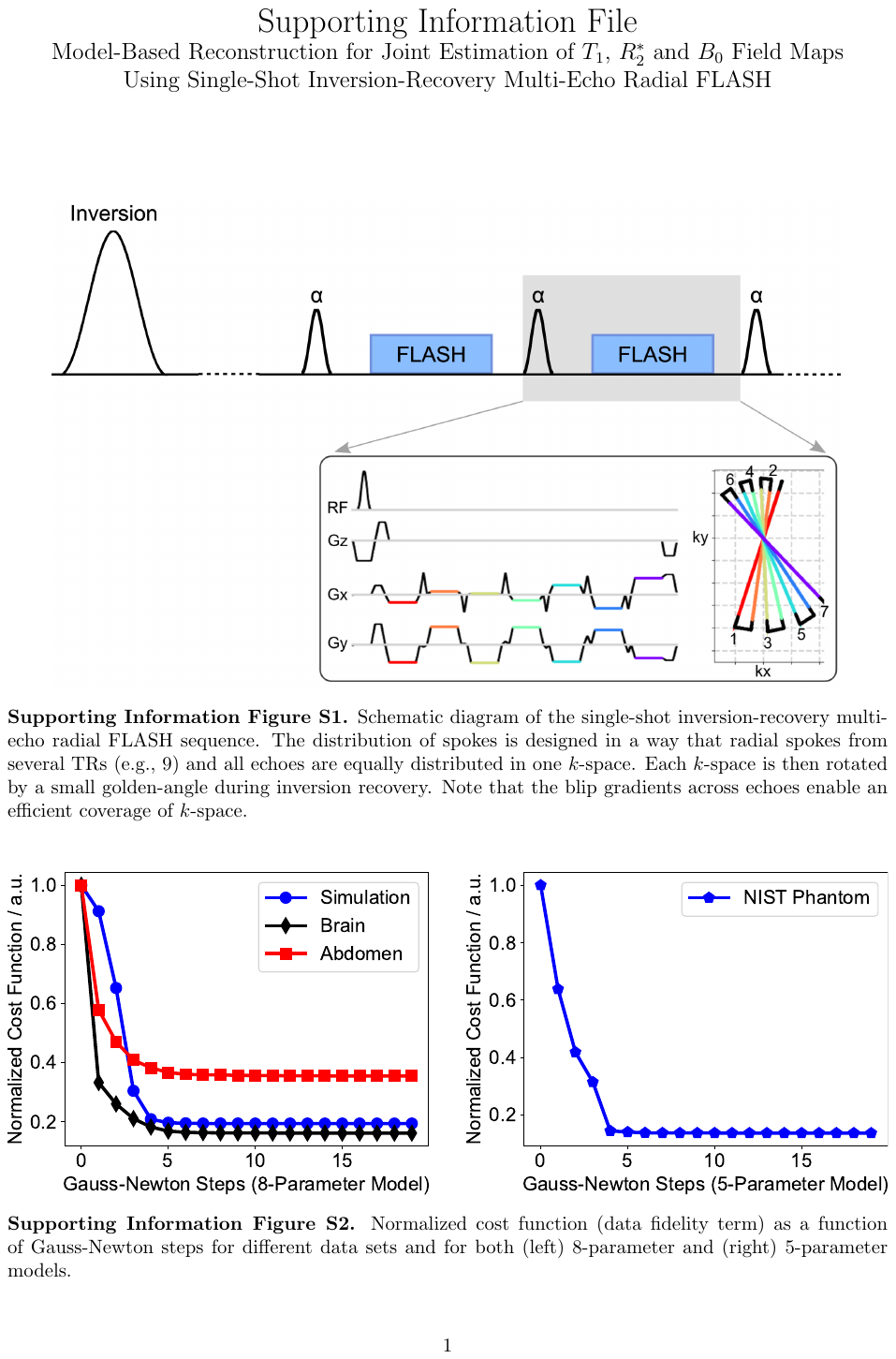}
%}

\end{document}